# A Method for Evaluating the Interpretability of Machine Learning Models in Predicting Bond Default Risk Based on LIME and SHAP

**Yan Zhang**[1], **Lin Chen**[1*], **Yixiang Tian**[1],

[1]School of Management and Economics, University of Electronic Science and Technology of China

*chenlin2@uestc.edu.cn

**Abstract**: Interpretability analysis methods for artificial intelligence models, such as LIME and SHAP, are widely used, though they primarily serve as post-model for analyzing model outputs. While it is commonly believed that the transparency and interpretability of AI models diminish as their complexity increases, currently there is no standardized method for assessing the inherent interpretability of the models themselves. This paper uses bond market default prediction as a case study, applying commonly used machine learning algorithms within AI models. First, the classification performance of these algorithms in default prediction is evaluated. Then, leveraging LIME and SHAP to assess the contribution of sample features to prediction outcomes, the paper proposes a novel method for evaluating the interpretability of the models themselves. The results of this analysis are consistent with the intuitive understanding and logical expectations regarding the interpretability of these models.

## 1 Introduction

In recent years, the substantial increase in computational power has fueled the rapid development of artificial intelligence algorithms, such as artificial neural networks and machine learning, which are now widely used across various fields. This progress has also driven advances in the concepts, methods, theoretical research, and practical applications of explainable artificial intelligence (XAI). However, to date, AI has not yet established a unified and authoritative conceptual framework. The current research issues and advancements in XAI can be summarized in the following domains.

The first domain is the concept and logical system of XAI [1]. This is a necessary theoretical foundation for a new field. Representative research such as Michael, William and Michael (2004) proposed the structure and concept of explainable artificial intelligence: explainable artificial intelligence can present an easy-to-understand reasoning chain to the user: from the user's command, through the knowledge and reasoning of the artificial intelligence, to the final decision [2]. Phillips and Hahn et al. (2021) summarized the four basic principles of explainable artificial intelligence[3]: "The system provides relevant evidence of input/output; the system provides an explanation that the user can understand; the explanation can accurately reflect the relationship between input and output; the system will only run when its output reaches sufficient confidence."

The second domain is explainable methods and theoretical research on artificial intelligence Currently, there are two mains widely accepted and applied "black box" model interpretability methods: LIME (Local Interpretable Model-agnostic Explanations) [4] and SHAP (Shapley Additive exPlanations) method [5]. LIME is a local interpretability method that maps an interpretable model

(such as linear regression) to each predicted value, enabling it to explain the predictions of complex models at that data point. The SHAP method is based on the Shapley value in game theory. It calculates a contribution score to the result for each feature and interprets the output results from the perspective of evaluating the importance of the feature.

The third domain is research on the application of explainable artificial intelligence in various industries. Currently, machine learning-related algorithms are mainly used as carriers to study the application of explainable artificial intelligence in geological science [6], wireless communications [7], recommendation systems [8], education [9], engineering technology [10], computer software [11], medicine [12], financial markets [13] and other fields.

The application of artificial intelligence (AI) algorithms in credit risk management has been a hot research topic in the financial field in recent years. With the development of machine learning, deep learning and other technologies, traditional credit risk assessment methods (such as rule-based scorecards or statistical models) are gradually replaced by data-driven algorithm models. These new methods not only improve prediction accuracy, but also can handle more complex and larger-scale data. Although these models provide higher accuracy, they lack interpretability, which can lead to opaque decision-making processes, especially in loan approvals, where users and regulators may be skeptical of the model's predictions. Therefore, the artificial intelligence interpretability analysis methods LIME and SHAP have been used to analyze the interpretability of artificial intelligence algorithms such as random forest, decision tree, XGBoost method, LightGBM, Neural network, and deep learning in credit risk assessment [15-23], including analyzing the key factors affecting default risk.

Given the lack of standardized evaluation methods for assessing the interpretability of artificial intelligence algorithms, this paper proposes a novel approach that integrates the LIME and SHAP methods to evaluate the interpretability of machine learning models. We hypothesize that, when a model demonstrates strong interpretability, the evaluation results of the sample features obtained through the LIME and SHAP methods should be consistent. Building on this premise, we utilize the correlation between the evaluation results from LIME and SHAP to quantify the interpretability of machine learning algorithms. This approach is applied to several models, including Random Forest (RF), Logistic Regression (LR), Decision Tree (DT), eXtreme Gradient Boosting (XGBoost).

The remainder of this paper is organized as follows: Section 2 provides an overview of several machine learning algorithms and their corresponding interpretability measures. Section 3 presents an application analysis using corporate bond data. Section 4 discusses the research methods and presents the conclusions, while Section 5 offers a summary of the paper.

## 2 Methods

### 2.1 LIME and SHAP

The LIME method is a technique used to interpret the predictions of machine learning models [4]. The core idea behind LIME is to provide local explanations for individual predictions made by a complex model, making the model's decision-making process more interpretable without requiring access to the internal structure of the model itself. LIME is a model-agnostic approach that generates local surrogate models to explain the behavior of a black-box model in the vicinity of a specific prediction. It works by perturbing the input data around a given instance, creating a set of modified data points, and then training an interpretable, simpler model (like a linear regression or decision tree) on this locally generated data. The surrogate model is then used to approximate the black-box model's behavior for that instance, providing insight into the factors that influenced the prediction.

The explanations provided by LIME for each observation x is obtained as follows [4]:

$$\xi(x) = \underset{g \in G}{\operatorname{argmin}} \mathcal{L}(f, g, \pi_x) + \Omega(g) \tag{1}$$

$G$ is a class of potentially interpretable models, $g \in G$ is An explanation considered as a model, and $f$ is the model being explained, $\pi_x$ is the proximity measure of an instance from $x$. $\Omega(g)$ is a measure of complexity. Because $\mathcal{L}(f, g, \pi_x)$ is the measure of how unfaithful $g$ is in approximating f in the locality defined by $\pi_x$, Then The goal is to minimize $\mathcal{L}(f, g, \pi_x)$ while having $\Omega(g)$ below enough to be interpretable by humans.

LIME is particularly useful in scenarios where a model is highly complex (e.g., deep learning, random forests, etc.), and the goal is to explain specific predictions in a way that is understandable to human users, such as in applications of healthcare, finance, and law, where interpretability is crucial.In summary, LIME enables the explanation of complex machine learning models by providing human-understandable, local approximations that shed light on the reasoning behind individual predictions.

The SHAP method in XAI used to explain the output of machine learning models by assigning each feature a contribution value [5]. It is grounded in Shapley values, which come from cooperative game theory. this way, the contribution of each explanatory variable to each point prediction can be assessed regardless of the underlying model. The idea behind SHAP is to fairly allocate the "credit" or "blame" for a model's prediction among its input features. It treats the machine learning model as a "game" where the input features (players) collaborate to produce a model's output (the "reward"). The SHAP method calculates how much each feature contributes to the difference between the model's prediction for a given instance and the average prediction across all instances. The Shapley value for a feature iii in a model prediction is computed as:

$$\phi_i(f) = \sum_{S \subseteq N \setminus \{i\}} \frac{|S|!(|N|-|S|-1)!}{|N|!} [f(S \cup \{i\}) - f(S)] \tag{2}$$

In Eq (2),
$\phi_i(f)$ is the Shapley value for feature $i$.
$S$ represents subsets of features excluding $i$.
$N$ is the set of all features.

$f(S)$ is the model's prediction for the subset of features $S$.

By analyzing SHAP values, it's possible to identify which features are most important to the model's predictions. SHAP values have become one of the most popular methods for model explain ability due to their theoretical grounding, consistency, and ease of implementation across various machine learning models.

**2.2 Several machine learning models**

We will first use the LIME and SHAP methods to analyze the interpretability of the following four algorithms for bond default prediction, and then measure the interpretability of the model itself based on the method proposed below.

(1) Logistic Regression (LR), it is a statistical method used for binary classification tasks, although it can be extended to multi-class classification problems as well. The core idea is to model the probability that a given input belongs to a certain class. It is based on the logistic (sigmoid) function, which maps any input into a value between 0 and 1, representing the probability of a given instance belonging to a particular class.

(2) Decision Tree (DT), it is a supervised machine learning algorithm used for both classification and regression tasks. It models decisions and their possible consequences in a tree-like structure, with nodes representing decisions or features, and branches representing outcomes. The goal of a decision tree is to partition the dataset into subsets based on the input features, eventually leading to a prediction for each instance. Decision Trees are a powerful, easy-to-understand model that can be used for both classification and regression tasks. They provide an intuitive way to split data based on feature values, making them particularly useful in real-world applications where interpretability is crucial. However, decision trees can suffer from overfitting, which is why they are often used in combination with techniques like pruning or ensemble methods such as Random Forests.

(3) Random Forest (RF), it is an ensemble learning method, particularly used for classification and regression tasks. It works by constructing a multitude of decision trees during training and then outputting the mode (classification) or mean (regression) of the individual tree predictions. The key strength of Random Forest lies in its ability to reduce overfitting by averaging multiple decision trees, which helps improve accuracy and generalization.

(4) Extreme Gradient Boosting (XGBoost), It is a popular and highly efficient machine learning algorithm based on the gradient boosting framework. It has gained widespread adoption due to its speed, accuracy, and performance in solving supervised learning problems, particularly in classification and regression tasks. XGBoost belongs to a family of ensemble methods, where multiple weak learners (usually decision trees) are combined to create a strong learner.The key idea behind gradient boosting is to build models sequentially, where each new model corrects the errors (residuals) made by the previous ones.

## 2.3 Measurement of interpretability of machine learning models

LIME and SHAP may all have an explanation model that is a linear function of binary variables [5]:

$$g(z') = \phi_0 + \sum_{i=1}^{M} \phi_i z_i' \quad (3)$$

In Eq(3), M is the number of simplified input features, $z_i' \in \{0,1\}^M$, $\phi_i \in R$ is an effect by each feature if $z_i' = 1$. In SHAP method, $\phi_i$ could be regarded as the Shapley value for feature $i$. On the other hand, in LIME method, $\phi_i$ could be regarded as the contribution of a feature variable in a linear regression model.

Let $z=[z_1', z_2', \cdots, z_M']$ is the Shapley value for feature variable, $w=[w_1', w_2', \cdots, w_M']$ is the contribution of a feature variable by LIME method, we define the Cosine Similarity of z and w is the measurement of interpretability of AI models (MIAI):

$$MIAI = \frac{z \cdot w}{\|z\| \|w\|} \quad (4)$$

Currently, there is a prevailing assumption that the interpretability of machine learning models is inversely related to the complexity of the algorithms employed. Linear models are generally considered the most interpretable, followed by decision trees and related models, while more complex algorithms, such as gradient boosting trees (e.g., XGBoost), are often viewed as less interpretable. However, there is a lack of a clear, standardized method or metric to quantitatively assess the interpretability of models. In light of this, we propose using the variable MIAI, as defined in Equation (4), as a measure of artificial intelligence model interpretability. The rationale behind this is that, for a model to be interpretable, the results of eigenvalue analyses conducted via the LIME and SHAP methods should exhibit a high degree of consistency, particularly in terms of the direction of influence. Therefore, we suggest using the correlation between LIME and SHAP results to assess the interpretability of AI models.

## 3 The Interpretability of Machine Learning Models

### 3.1 Sample data and definition of variables

Financial data for a total of 6,471 bond issuers in 2018 were obtained from the Wind database, with 50 of these issuers defaulting in 2019. Based on existing literature, this study selected 16 financial indicators derived from an analysis of four key aspects of bond issuers: profitability, operational capacity, solvency, and capital structure (see Table 1), in addition to the external audit opinion. External audit opinions are classified into four main types: (1) unable to express an opinion, (2) unqualified opinion with emphasis of matter paragraph, (3) qualified opinion, and (4) standard unqualified opinion. A numerical value is assigned to each opinion type to reflect the degree to which the financial statements are deemed acceptable: 1 for 'unable to express an opinion,' 2 for 'unqualified opinion with emphasis of matter paragraph,' 3 for 'qualified opinion,' and 4 for 'standard unqualified opinion. According to the general classification size of training samples and testing samples, we divide 80% of the total samples into training samples and the remaining 20% into testing samples.

Table1. Definition of variables

| Variable symbols | | Definition of variables |
|---|---|---|
| Identifying variables | Default | if bond default, then default=1, or default =0 |
| Feature variables | Prcb | Profit rate of core businesses |
| | Igrb | Income growth rate of core businesses |
| | Roa | Return on asset |
| | Roe | Return on equity |
| | EBII | EBITDA/Total Income |
| | Ocebi | Operating cash/EBITDA |
| | Lr | Liquidity ratio |
| | Qr | Quick ratio |
| | Rst | Rate of stock turnover |
| | Alr | Asset-liability ratio |
| | Sdtd | Short-term debt / total debt |
| | Ditc | Debt with interest /total investment capita |
| | Mtd | Monetary /total  debt |
| | Im | Interest multiples |
| | Ebit | EBITDA / total debt with interest |
| | Aou | Audit opinion |

## 3.2 Prediction by machine learning algorithm

Firstly, using the algorithmic functions provided by Python, we trained models for four methods: Random Forest (RF), Logistic Regression (LR), Decision Tree (DT), and eXtreme Gradient Boosting (XGBoost), and evaluated their performance on the test dataset. The models were assessed using the Area Under the Curve (AUC), a common metric for model evaluation. The performance of each model is summarized in Table 2. Given the presence of class imbalance in the dataset, accuracy was not employed as a performance metric. To ensure the integrity of the interpretability analysis, we retained the original, imbalanced dataset without applying any data balancing techniques.

Table.2 Comparative prediction of several models

| | LR | DT | RF | XGBoost |
|---|---|---|---|---|
| AUC | 0.6368 | 0.9372 | 0.9917 | 0.9911 |

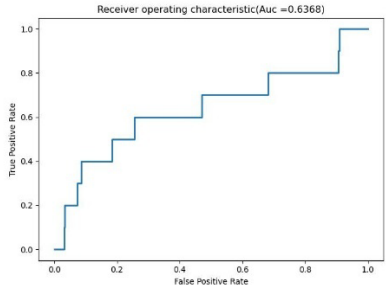
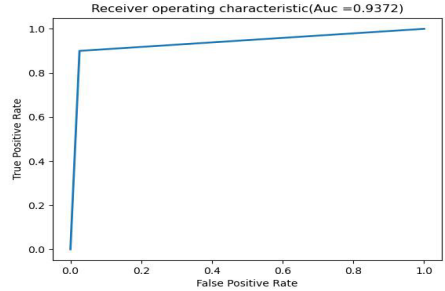

Figure.1 AUC of Logistic Regression     Figure.2 AUC of Decision Tree

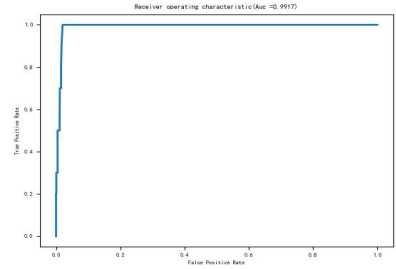
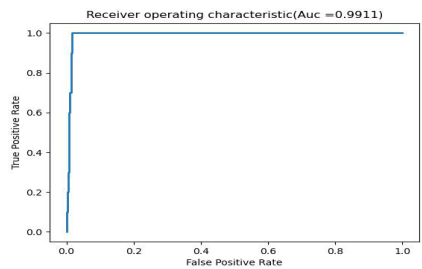

Figure.3 AUC of Random Forest     Figure.4 AUC of XGBoost

Regarding the performance of the predictive models, it is evident that XGBoost and Random Forest outperform Logistic Regression (LR) and Decision Tree (DT) algorithms in terms of classification accuracy. Specifically, the LR model demonstrates the poorest performance on this imbalanced dataset, a result that aligns with common expectations regarding the relative effectiveness of these algorithms under such conditions.

### 3.3 Measurement of interpretability based on LIME and SHAP

On our test dataset, we employed both SHAP and LIME methods to analyze the interpretability of the various models. Given that the LIME algorithm provides explanations at the individual sample, we applied LIME to assess the impact of each feature value on the prediction result for each sample point. Subsequently, we computed the average impact across all test samples to obtain an overall explanation of how each feature influences the model's output, as presented in Table 3.

Table.3 The impact of features on model output (LIME)

| Feature variables | LR | DT | RF | XGBoost |
|---|---|---|---|---|
| Prcb | 0.0006 | -0.0042 | -0.00982 | -0.0638 |
| Igrb | 0.0679 | -0.0002 | -0.0059 | -0.0080 |
| Roa | -0.0111 | -0.0155 | -0.02057 | -0.0402 |
| Roe | -0.0645 | 0.0125 | -0.01548 | -0.0070 |
| EBII | 0.0015 | 0.0009 | -0.00645 | 0.0014 |
| Ocebi | 0.0013 | 0.0008 | -0.00873 | -0.0012 |
| Lr | 0.0707 | 0.0002 | -0.00237 | 0.0184 |
| Qr | 0.0445 | -0.0036 | -0.00337 | -0.0007 |
| Rst | 0.0274 | -0.0011 | 0.000827 | -0.0235 |
| Alr | 0.0091 | 0.0002 | -0.00498 | 0.0259 |
| Sdtd | 0.0043 | -0.0001 | -0.00071 | -0.0654 |
| Ditc | -0.0157 | -0.0004 | -0.01196 | -0.0545 |
| Mtd | 0.0023 | -0.0002 | -0.00846 | -0.0118 |
| Im | 0.0484 | 0.0001 | 0.005066 | -0.0378 |
| Ebit | 0.0763 | 0.0002 | -0.00662 | -0.0136 |
| Aou | -0.0189 | 0.0000 | 0.0000 | -0.1348 |

We also calculate the average effect of the SHAP model on each feature value as presented in Table 4.

Table.4 The impact of features on model output (SHAP)

| Feature variables | LR | DT | RF | XGBoost |
|---|---|---|---|---|
| Prcb | 0.0010 | 0.0035 | 0.0006 | -0.0116 |
| Igrb | 23.4839 | 0.0011 | 0.0001 | -0.0016 |
| Roa | 0.0875 | -0.0086 | -0.0025 | -0.0038 |
| Roe | 0.0245 | -0.0015 | 0.0017 | -0.0324 |
| EBII | 0.0001 | -0.0005 | 0.0004 | 0.0021 |
| Ocebi | 0.0001 | 0.0023 | 0.0004 | -0.0010 |
| Lr | -3.7279 | 0.0009 | -0.0001 | 0.0026 |
| Qr | -1.1489 | 0.0000 | 0.0005 | -0.0007 |
| Rst | -0.2269 | 0.0004 | 0.0000 | 0.0354 |
| Alr | -0.0235 | 0.0000 | 0.0005 | 0.0003 |
| Sdtd | 0.1384 | 0.0001 | 0.0003 | 0.0091 |
| Ditc | 0.0018 | -0.0002 | -0.0006 | -0.0011 |
| Mtd | 0.0123 | 0.0003 | -0.0004 | -0.0060 |
| Im | -0.1295 | 0.0001 | 0.0007 | 0.0017 |
| Ebit | 0.6573 | 0.0000 | 0.0002 | 0.0003 |
| Aou | -0.0119 | 0.0267 | 0.0025 | 0.0000 |

Fig.5-Fig.8 shows the summary plot. It helps us overview which features are most important for four model

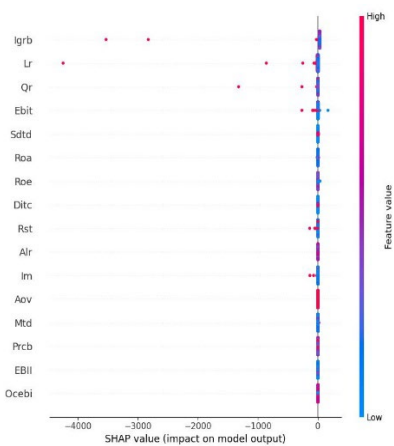 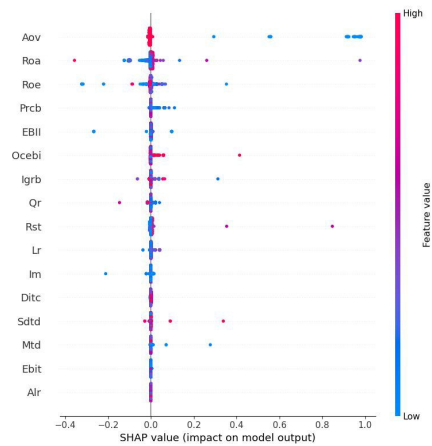

Figure.5 SHAP value of LR model        Figure.6 SHAP value of DT model

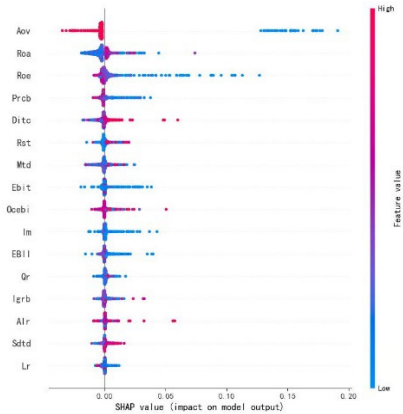 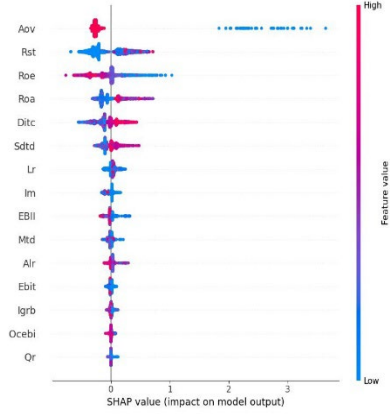

Figure.7 SHAP value of RF model    Figure.8 SHAP value of XGBoost model

Finally, based on the definition provided in Equation (4), we computed the correlation of the contributions of each feature, as determined by the LIME and SHAP methods, to assess the interpretability of the four models. The results of this analysis are presented in Table 5 and Figure 9

Table.5 The measurement of interpretability of model

|  | LR | DT | RF | XGBoost |
|---|---|---|---|---|
| MIAI | 0.3459 | 0.1708 | 0.1430 | -0.0182 |

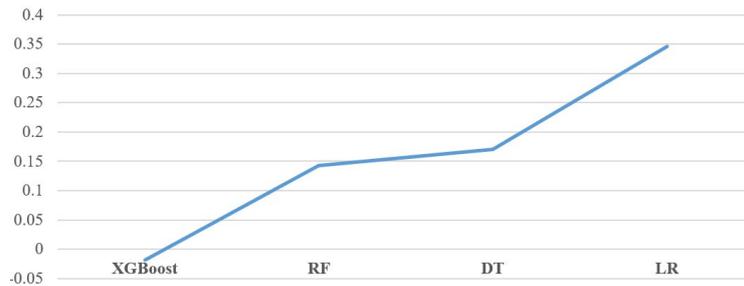

Figure.9 Measurement of interpretability of four models

## 4 Discussion

According to conventional corporate finance theory, ceteris paribus, certain indicators are negatively associated with default risk, meaning that larger values of these indicators correspond to a lower likelihood of default. Conversely, the remaining three indicators are positively related to default risk, such that higher values of these indicators are associated with an increased probability of default. For a model to demonstrate strong explanatory power, the analysis of feature contributions should align with established financial theory.

Table.6　Comparative analysis with financial theory（LIME）

| Feature variables | The impact of on default based on financial theory | LR | DT | RF | XGBoost |
|---|---|---|---|---|---|
| Prcb | - | ＋ | - | - | - |
| Igrb | - | ＋ | - | - | - |
| Roa | - | - | - | - | - |

| Feature variables | The impact of on default based on financial theory | LR | DT | RF | XGBoost |
|---|---|---|---|---|---|
| Roe | - | - | + | - | - |
| EBII | - | + | + | - | + |
| Ocebi | - | + | + | - | - |
| Lr | - | + | + | - | + |
| Qr | - | + | - | - | - |
| Rst | - | + | - | + | - |
| Alr | + | + | + | - | + |
| Sdtd | + | + | - | - | - |
| Ditc | + | - | - | - | - |
| Mtd | - | + | - | - | - |
| Im | - | + | + | + | - |
| Ebit | - | + | + | - | - |
| Aou | - | - | + | + | - |

Table.7　Comparative analysis with financial theory（SHAP）)

| Feature variables | The impact of on default based on financial theory | LR | DT | RF | XGBoost |
|---|---|---|---|---|---|
| Prcb | - | + | + | + | - |
| Igrb | - | + | + | + | - |
| Roa | - | + | - | - | - |
| Roe | - | + | - | + | - |
| EBII | - | + | - | + | + |
| Ocebi | - | + | + | + | - |
| Lr | - | - | + | - | + |
| Qr | - | - | + | + | - |
| Rst | - | - | + | + | + |
| Alr | + | - | + | + | + |
| Sdtd | + | + | + | + | + |
| Ditc | + | + | - | - | - |
| Mtd | - | + | + | - | - |
| Im | - | - | + | + | + |
| Ebit | - | + | + | + | + |
| Aou | - | - | + | + | + |

　　From Tables 6 and 7, it is evident that although the simple linear model does not yield high accuracy, both the LIME and SHAP methods exhibit strong consistency in determining the direction of the impact of features on default risk. While this consistency may not always align with financial theory, it nevertheless demonstrates the stability of the model with respect to the specific dataset, and is consistent with traditional interpretations of feature contributions. In our analysis, the LR model shows consistent results between LIME and SHAP for 9 features, while the DT model shows consistency for 8 features. In contrast, the RF and XGBoost models exhibit consistency for only 3 features. Accordingly, based on the metric defined in Equation (4), XGBoost demonstrates the lowest interpretability, whereas the LR model shows the highest interpretability.

Additionally, in the interpretability analysis of the four models using the LIME method, the assessment of Return on Assets (ROA) is both consistent and aligns with financial theory. In contrast, the SHAP method reveals that the evaluation of the short-term debt to total debt ratio (Sdtd) is consistent with financial theory. These findings suggest that the profitability of corporate assets (as measured by ROA) and the level of short-term debt are key factors influencing corporate bond defaults.

## 5 Conclusion

This paper aims to propose a quantitative indicator for measuring the interpretability of explainable artificial intelligence (XAI) models. Specifically, we suggest assessing the interpretability of machine learning algorithms through a combination of LIME and SHAP. Using our dataset, we evaluated the interpretability of four commonly employed classification algorithms, and the results align with intuitive expectations.

Given that the interpretability of AI models remains a subject of ongoing debate, our proposed method may have certain limitations. For instance, as the foundational LIME and SHAP methods are not entirely robust across all data environments, different conclusions might be reached when applying our method to varied datasets. Nonetheless, we believe the approach we have introduced is a valid contribution to addressing this issue, and we foresee future research that could integrate additional post-hoc interpretability methods to further refine and enhance the evaluation of model interpretability.